\documentclass[superscriptaddress,twocolumn, showpacs,amsmath,amssymb,aps,prl,floatfix, 10pt]{revtex4-2}
\usepackage[latin1]{inputenc}
\usepackage[american]{babel}
\usepackage[T1]{fontenc}
\usepackage{mathrsfs}
\usepackage{hyperref}
\usepackage{multirow}
\usepackage{bm, epsfig}
\usepackage{graphicx}
\usepackage{subeqnarray}
\usepackage{bbold}
\usepackage{upgreek}
\usepackage{tikz}
\usepackage{slashed}
\usepackage{lineno}
\usepackage{longtable}
\usepackage{booktabs}
\usepackage{float}
\usepackage{verbatim}

\hypersetup{colorlinks=true, urlcolor=blue, citecolor=blue, filecolor=blue, linkcolor=blue}

\begin{document}

\title{Hadronic vacuum polarization effect in muonic atoms}

\date{\today}

\author{Zoia~Mandrykina}
\email[Email: ]{zoia.mandrykina@mpi-hd.mpg.de} 
\affiliation{Max-Planck-Institut f\"{u}r Kernphysik, Saupfercheckweg 1, 69117 Heidelberg, Germany}
\author{Moritz~Thierfeld}
\affiliation{Max-Planck-Institut f\"{u}r Kernphysik, Saupfercheckweg 1, 69117 Heidelberg, Germany}
\author{Natalia~S.~Oreshkina}
\email[Email: ]{Natalia.Oreshkina@mpi-hd.mpg.de} 
\affiliation{Max-Planck-Institut f\"{u}r Kernphysik, Saupfercheckweg 1, 69117 Heidelberg, Germany}

\begin{abstract}
The hadronic vacuum polarization (HVP) corrections to energy levels in muonic atoms are studied systematically across the periodic table. 
Two nuclear charge distribution models have been considered, with partly analytical solution for an homogeneously charged sphere model and direct numerical integration for a Fermi nuclear charge distribution.
Our comparison reveals a good agreement between both approaches with differences typically below a few percent.
The magnitude of HVP corrections increases with nuclear charge, displaying irregularities that correlate directly with the non-monotonic behavior of nuclear radii.
Despite strong relativistic effects at high $Z$, the ratio of hadronic to muonic vacuum polarization remains remarkably stable, deviating by less than $10\%$ from the established $67\%$ non-relativistic benchmark value.
\end{abstract}

\maketitle


\section{Introduction} 

Precision spectroscopy of hydrogen-like and muonic systems provides a robust platform for testing QED and probing nuclear structure~\cite{Knecht2020, 2210.16929, PhysRevC.111.034313}. QED has proven to be one of the most precisely validated theories in physics, with remarkable agreement between theory and experiment for atomic systems ~\cite{h-like-tin, PhysRevLett.130.173001}. As experimental techniques continue to advance, increasingly subtle quantum effects must be accounted for to maintain a corresponding theoretical accuracy \cite{Antognini}.

While electronic systems serve as excellent benchmarks for self-energy correction, vacuum polarization~(VP) effects dominate in muonic systems due to the muon's significantly larger mass and closer proximity to the nucleus~\cite{Karshenboim2006,RevModPhys.96.015001, BORIE2012733}. These factors enhance the sensitivity of muonic systems to short-ranged quantum effects~\cite{PhysRevC.1.1176,PhysRevA.109.042811}, including hadronic vacuum polarization (HVP), which involves virtual hadronic particles that interact via the strong force. 
In muonic atoms, the reduced Bohr radius significantly enhances HVP contributions, which become a measurable fraction of observables such as the Lamb shift and fine-structure splitting~\cite{PhysRevA.18.324}. The calculation of HVP effects needs more complex theoretical approaches compared to those required for electronic or muonic VP due to the non-perturbative nature of quantum chromodynamics at low energies.
High-$Z$ systems, in particular, exhibit enhanced relativistic and nuclear structure effects ~\cite{PhysRevA.108.052824, PhysRevResearch.4.L042040, RevModPhys.54.67}, further amplifying the impact of HVP correction.

Recently, Breidenbach et al.~\cite{Breidenbach_PhysRevA.106.042805} have evaluated HVP in a semiempirical way for hydrogenlike ions and muonic hydrogen.
In our work, we extend this methodology to systematically compute HVP corrections across a wide range of nuclear charges $Z$ in muonic atoms. 
We adopt a semi-empirical, dispersion-relation-based approach, using a parametrized hadronic polarization function fitted to experimental $e^+e^- \rightarrow$ hadrons cross section data~\cite{BURKHARDT1989}. The resulting potential is used in fully relativistic energy shift calculations, including finite nuclear size effects with the two nuclear charge distribution models: well-founded 2-parameter Fermi nuclear charge distribution and simpler but less realistic homogeneously charged sphere model. Our results provide improved theoretical predictions for low-lying energy levels in muonic systems compared to \cite{PhysRevD.11.230} and offer insight into the scaling behaviour of HVP concerning nuclear charge and structure. 

Relativistic muonic units ($m_\mu = \hbar = c = 1$) and the Heaviside charge units ($e^2 = 4\pi\alpha$) are used throughout the paper.
\section{Theory}
While the leptonic polarization function has an analytical expression, the hadronic case requires a semi-empirical construction of its regularized real part due to the non-perturbative behavior of the strong interaction:
\begin{equation}
    {\rm Re}[\Pi^{\rm R}_{ \rm Had}(q^2)] = A_i + B_i \ln(1 + C_i|q^2|),
\end{equation}
with $A_i$, $B_i$, $C_i$ defined piecewise for different energy intervals of $q^2$. The parameters corresponding to the low-energy domain ($0.0~\text{--}~0.7~\mathrm{GeV}$) are particularly relevant in atomic physics calculations due to the comparable energy scale.

In this work, we restrict the analysis to the low-energy domain, which corresponds to the first parameterization region. The adopted values of the fit parameters are \( A_1 = 0 \), \( B_1 = 0.0023092 \), and \( C_1 = 3.9925370~\mathrm{GeV}^{-2} \)~\cite{BURKHARDT2001}. With this choice, the hadronic Uehling potential simplifies to
\begin{multline}
    V_{\rm HVP}(r) = -\frac{2e}{\pi}
    \int_0^{0.7~\mathrm{GeV}} dq~j_0(qr) \tilde{\rho}(q) \\
    \times B_1\ln(1 + C_1 q^2),
    \label{begin}
\end{multline}
where \( j_0(qr) \) is the spherical Bessel function and \( \tilde{\rho}(q) \) is Fourier transformation of the nuclear charge distribution $\rho(x)$ in coordinate space. 

This expression can be further simplified using analytical integrals and represented in an easier form for computational purposes:
\begin{multline}
    V_{\rm HVP}(r) = -\frac{4\pi e B_1 \sqrt{C_1}}{r}\int^\infty_{0} dx \, x \rho(x) \\
    \times
    \left[
    E_2\left(\frac{|r - x|}{\sqrt{C_1}}\right) - E_2\left(\frac{r + x}{\sqrt{C_1}}\right)
    \right],
    \label{HVP}
\end{multline}
the exponential integral being
\begin{equation}
    E_n(x) = \int_1^\infty dt\frac{e^{-x t}}{t^n}.
\end{equation}

In our work, we have considered two different nuclear charge distribution models to analyze the model dependence of HVP contribution. 
The first one is the well-known 2-parameter Fermi model, where a direct numerical integration of Eq.~(\ref{begin}) has been used. 
Another model is the less realistic but more simple homogeneous charged sphere model where the corresponding radius is related to the root-mean-square (rms) nuclear charge radius through $R = \sqrt{5/3\langle r^2\rangle}$.

In the case of the spherical charge distribution, integral (\ref{HVP}) can be further simplified and represented as a sum 
for two different regions, outside the nucleus ($r$ \textgreater $R$):
\begin{multline}
V^{\rm sphere}_{\rm HVP, out}(r) = -\frac{3Z\alpha B_1 \sqrt{C_1}}{r R^3} \\
\times
\biggl\{
\sqrt{C_1}R \left[  E_3\left(\frac{|r - R|}{\sqrt{C_1}}\right) + E_3\left(\frac{r + R}{\sqrt{C_1}}\right) \right] \\
-
C_1 \left[ E_4\left(\frac{|r - R|}{\sqrt{C_1}}\right) - E_4\left(\frac{r + R}{\sqrt{C_1}}\right) \right] 
\biggr\},
\label{6}
\end{multline}
and inside it $r \leqslant R$:
\begin{multline}
V^{\rm sphere}_{\rm HVP, in}(r) = -\frac{3Z\alpha B_1 \sqrt{C_1}}{r R^3} \\
\times
\biggl\{
\sqrt{C_1}r + \sqrt{C_1}R E_3\left(\frac{r + R}{\sqrt{C_1}}\right) + C_1 E_4\left(\frac{r + R}{\sqrt{C_1}}\right) \\
-\frac{1}{6}e^{\frac{r-R}{\sqrt{C_1}}}
\left[ 2C_1+\sqrt{C_1}(r+2R)+(r-R)(r+2R)\right] \\
-\frac{(r-R)^2(r+2R)}{6\sqrt{C_1}}E_1\left(\frac{R - r}{\sqrt{C_1}}\right)
\biggr\}.
\label{7}
\end{multline}

The energy level correction in the first order of the perturbation theory for the bound muon state with the main quantum number $n$, the Dirac quantum number $\kappa$ and the projection of the total angular momentum $m$,  can be represented as 
\begin{multline}
    \Delta E_{n\kappa m} =
    \langle n\kappa m | V |n\kappa m \rangle
    \\
    = \int_0^\infty dr~V(r)
    \left(G^2_{n\kappa}(r) + F^2_{n\kappa}(r)
    \right),
\end{multline}
where $V$  is 
{defined as described above depending on the nuclear model used.}

\section{Results and discussion}
\begin{figure*}[ht!]
    \centering
    \includegraphics[width=0.95\textwidth]{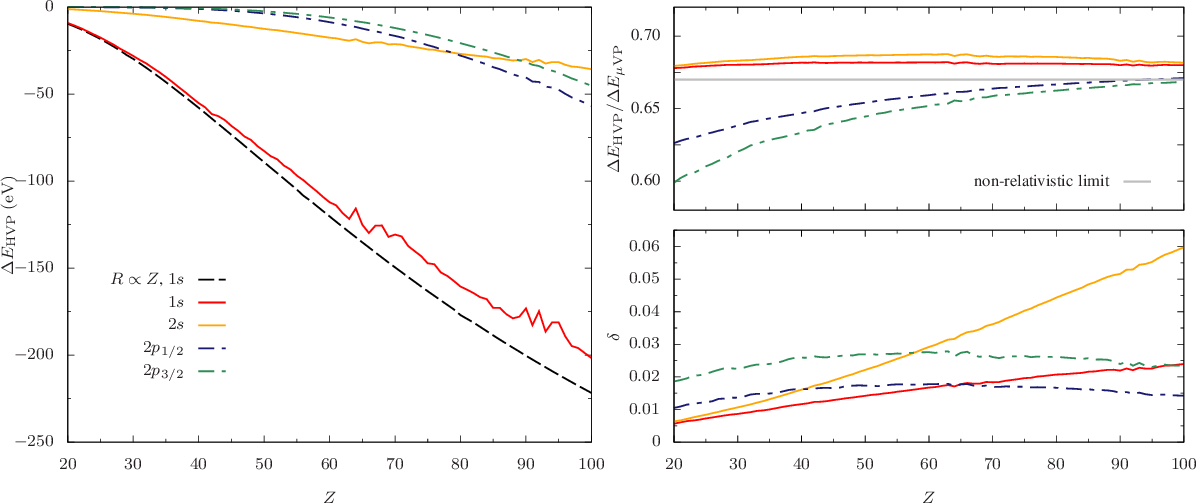}
    \caption{Left panel: Hadronic vacuum polarization (HVP) energy shift for different muonic states of the H-like ions. Right top panel: Ratio of the HVP energy shift to Uehling energy correction. The right bottom panel: Relative difference between the 2-parameter Fermi and the homogeneous charged sphere nuclear charge distribution models defined by Eq.~(\ref{diffdif}). All quantities are shown as functions of nuclear charge $Z$.}
    \label{fig:combined}
\end{figure*}
The results for the HVP correction to the energies of the $1s_{1/2}$, $2s_{1/2}$, $2p_{1/2}$, and $2p_{3/2}$ muonic states across a wide range of nuclear charge values $Z$ are presented in Fig.~\ref{fig:combined}.
For all four states, we observe a general monotonic increase in the magnitude of the energy shift with increasing $Z$; however in the high-$Z$ region ($Z \geq 60$) we observe a spiky, slightly irregular behavior, especially for the $1s$ state. 
Since these anomalies correlate with the non-monotonic behavior of nuclear charge radii~\cite{Angeli2013Table}, we repeated our calculations using a simple smooth approximation for the rms radius $R = 1.2\, Z^{1/3}$.
The corresponding results are shown in left panel of Fig.~\ref{fig:combined} with black dashed line. This simple correction restores the monotonic profile, confirming that such irregularities correspond to fluctuations of the rms values.

The $1s_{1/2}$ state exhibits, as expected, the highest sensitivity to the HVP correction, as its wave function is predominantly localized in close proximity to the nucleus, where the HVP potential has its maximum effect. For muonic atoms, this effect is also significantly enhanced compared to electronic systems due to the muon's larger mass. 

Next, we compare the HVP effect with the muonic vacuum polarization ($\mu$VP) correction. 
Since an order of magnitude in the hierarchy of VP corrections is defined by the particle mass of the corresponding particle-antiparticle in the loop, this comparison is both natural and instructive.
Additionally, by expressing the HVP correction as a fraction of the $\mu$VP term, we can better understand its relative significance within the overall QED correction framework and establish consistent scaling patterns across different atomic systems and energy levels.
Previously, this ratio was calculated only non-relativistically for the hydrogen case \cite{Friar}.
It was demonstrated that the HVP contribution amounts to approximately $67\%$ of the $\mu$VP shift. This widely accepted proportion has been used as a standard reference in the field, but its applicability to higher-$Z$ muonic systems remains unexplored. Our systematic investigation provides the first comprehensive analysis of how such relation changes with increasing nuclear charge, filling a significant gap in our understanding of vacuum polarization effects in exotic atoms.
The upper right panel of Fig.~\ref{fig:combined} reveals the $Z$-dependent evolution of the $\rm \Delta E_{\rm HVP}/\Delta E_{\rm \mu VP}$ ratio across the four lowest muonic states. All values lie within a narrow range $(0.60-0.70)$, confirming its stability across the periodic table while revealing important state-dependent relativistic effects.

The $s$-states maintain nearly identical and remarkably stable values around $0.675-0.68$ throughout the entire $Z$ range from $20$ to $100$. These states maintain the highest ratio values among all four considered states, consistently staying above the level of $67\%$.
At the same time, both $2p_{1/2}$ and $2p_{3/2}$ states show qualitatively different behaviors, beginning with lower values of $~0.62-0.63$ at $Z = 20$ and exhibiting a monotonic growth toward the non-relativistic benchmark value.
This demonstrates the importance of realistic nuclear modeling over simplified approximations, particularly in the low- and medium-$Z$ region.

The model dependence has been investigated by introducing a relative difference obtained by adopting a potential of a 2-parameter Fermi and of an homogeneous charged sphere:
\begin{equation}
\delta = \frac{\Delta E^{\rm Fermi}_{\text{HVP}} - \Delta E^{\rm sphere}_{\text{HVP}}}{\Delta E^{\rm Fermi}_{\text{HVP}}}.    
\label{diffdif}
\end{equation}
For most nuclei across the periodic table, the two approaches yield similar results, with differences typically below $0.03$ for the $1s$ state, confirming the reliability of both models. 
The calculations appear even more robust in the case of the $p$-states (blue and green dashed lines): the delta values stay constantly below 0.03 throughout the entire $Z$ range and showing minimal variation even for high $Z \gtrsim 70$. The nearly identical behavior of $2p_{1/2}$ and $2p_{3/2}$ states reflects their similar radial distributions and reduced nuclear penetration.

More significant discrepancies (up to 0.05-0.06) emerge only for heavy nuclei with $Z\gtrsim 90$ in $2s$ state.
These deviations suggest that the homogeneously charged sphere model may not fully adequately represent the nuclear charge distribution for these particular elements, especially when calculating the HVP correction.
In fact, the same behavior can also be observed for $\mu$VP correction, suggesting a general trend for highly-localized potentials.

\section{Conclusions}
We have calculated the hadronic vacuum polarization corrections to energy levels in muonic atoms across a wide range of nuclear charges from $Z = 20$ to $Z = 100$, covering the four states: $1s_{1/2}$, $2s_{1/2}$, $2p_{1/2}$, and $2p_{3/2}$. 
We verified the predicted scaling trends and found that while HVP corrections generally increase with nuclear charge following the expected theoretical behavior, pronounced deviations emerge in the high-$Z$ region that correlate directly with the non-monotonic behavior of experimental nuclear charge radii.

We studied the model dependence by comparing results from the 2-parameter Fermi nuclear charge distribution with those from the homogeneously charged sphere model, revealing excellent agreement for light and medium nuclei but significant discrepancies reaching up to $5-6\%$ for heavy elements.

For practical applications in high-precision spectroscopy of muonic atoms, this 
implies that the HVP correction becomes increasingly important alongside the 
$\mu$VP effect in heavy elements, where the HVP contribution reaches up to $68\%$ 
of $\mu$VP.
However, deviations from this trend can reach nearly $10\%$, highlighting the necessity of more rigorous calculations when high precision is required.

\section{Acknowledgements}
The authors would like to thank V.~A.~Yerokhin for fruitful discussions. This article comprises parts of the PhD thesis work of Z. M. to be submitted to Heidelberg University.

\bibliography{refs}

\begin{thebibliography}{21}%
\makeatletter
\providecommand \@ifxundefined [1]{%
 \@ifx{#1\undefined}
}%
\providecommand \@ifnum [1]{%
 \ifnum #1\expandafter \@firstoftwo
 \else \expandafter \@secondoftwo
 \fi
}%
\providecommand \@ifx [1]{%
 \ifx #1\expandafter \@firstoftwo
 \else \expandafter \@secondoftwo
 \fi
}%
\providecommand \natexlab [1]{#1}%
\providecommand \enquote  [1]{``#1''}%
\providecommand \bibnamefont  [1]{#1}%
\providecommand \bibfnamefont [1]{#1}%
\providecommand \citenamefont [1]{#1}%
\providecommand \href@noop [0]{\@secondoftwo}%
\providecommand \href [0]{\begingroup \@sanitize@url \@href}%
\providecommand \@href[1]{\@@startlink{#1}\@@href}%
\providecommand \@@href[1]{\endgroup#1\@@endlink}%
\providecommand \@sanitize@url [0]{\catcode `\\12\catcode `\$12\catcode
  `\&12\catcode `\#12\catcode `\^12\catcode `\_12\catcode `\%12\relax}%
\providecommand \@@startlink[1]{}%
\providecommand \@@endlink[0]{}%
\providecommand \url  [0]{\begingroup\@sanitize@url \@url }%
\providecommand \@url [1]{\endgroup\@href {#1}{\urlprefix }}%
\providecommand \urlprefix  [0]{URL }%
\providecommand \Eprint [0]{\href }%
\providecommand \doibase [0]{https://doi.org/}%
\providecommand \selectlanguage [0]{\@gobble}%
\providecommand \bibinfo  [0]{\@secondoftwo}%
\providecommand \bibfield  [0]{\@secondoftwo}%
\providecommand \translation [1]{[#1]}%
\providecommand \BibitemOpen [0]{}%
\providecommand \bibitemStop [0]{}%
\providecommand \bibitemNoStop [0]{.\EOS\space}%
\providecommand \EOS [0]{\spacefactor3000\relax}%
\providecommand \BibitemShut  [1]{\csname bibitem#1\endcsname}%
\let\auto@bib@innerbib\@empty
\bibitem [{\citenamefont {Knecht}\ \emph {et~al.}(2020)\citenamefont {Knecht},
  \citenamefont {Skawran},\ and\ \citenamefont {Vogiatzi}}]{Knecht2020}%
  \BibitemOpen
  \bibfield  {author} {\bibinfo {author} {\bibfnamefont {A.}~\bibnamefont
  {Knecht}}, \bibinfo {author} {\bibfnamefont {A.}~\bibnamefont {Skawran}},\
  and\ \bibinfo {author} {\bibfnamefont {S.}~\bibnamefont {Vogiatzi}},\
  }\bibfield  {title} {\bibinfo {title} {Study of nuclear properties with
  muonic atoms},\ }\bibfield  {journal} {\bibinfo  {journal} {Eur. Phys. J.
  Plus}\ }\textbf {\bibinfo {volume} {135}},\ \href
  {https://doi.org/https://doi.org/10.1140/epjp/s13360-020-00777-y}
  {https://doi.org/10.1140/epjp/s13360-020-00777-y} (\bibinfo {year}
  {2020})\BibitemShut {NoStop}%
\bibitem [{\citenamefont {Saito}\ \emph {et~al.}()\citenamefont {Saito},
  \citenamefont {Niikura}, \citenamefont {Matsuzaki},\ and\ \citenamefont
  {et~al.}}]{2210.16929}%
  \BibitemOpen
  \bibfield  {author} {\bibinfo {author} {\bibfnamefont {T.~Y.}\ \bibnamefont
  {Saito}}, \bibinfo {author} {\bibfnamefont {M.}~\bibnamefont {Niikura}},
  \bibinfo {author} {\bibfnamefont {T.}~\bibnamefont {Matsuzaki}},\ and\
  \bibinfo {author} {\bibnamefont {et~al.}},\ }\bibfield  {title} {\bibinfo
  {title} {Muonic-atom spectroscopy and impact on nuclear structure and
  precision {QED} theory}\ }\href
  {https://doi.org/https://doi.org/10.48550/arXiv.2210.16929}
  {https://doi.org/10.48550/arXiv.2210.16929}\BibitemShut {NoStop}%
\bibitem [{\citenamefont {Saito}\ \emph {et~al.}(2025)\citenamefont {Saito},
  \citenamefont {Niikura}, \citenamefont {Matsuzaki},\ and\ \citenamefont
  {et~al.}}]{PhysRevC.111.034313}%
  \BibitemOpen
  \bibfield  {author} {\bibinfo {author} {\bibfnamefont {T.~Y.}\ \bibnamefont
  {Saito}}, \bibinfo {author} {\bibfnamefont {M.}~\bibnamefont {Niikura}},
  \bibinfo {author} {\bibfnamefont {T.}~\bibnamefont {Matsuzaki}},\ and\
  \bibinfo {author} {\bibnamefont {et~al.}},\ }\bibfield  {title} {\bibinfo
  {title} {Muonic x-ray measurement for the nuclear charge distribution: The
  case of stable palladium isotopes},\ }\href
  {https://doi.org/10.1103/PhysRevC.111.034313} {\bibfield  {journal} {\bibinfo
   {journal} {Phys. Rev. C}\ }\textbf {\bibinfo {volume} {111}},\ \bibinfo
  {pages} {034313} (\bibinfo {year} {2025})}\BibitemShut {NoStop}%
\bibitem [{\citenamefont {Morgner}\ \emph {et~al.}(2023)\citenamefont
  {Morgner}, \citenamefont {Tu}, \citenamefont {K\"{o}nig},\ and\ \citenamefont
  {et~al.}}]{h-like-tin}%
  \BibitemOpen
  \bibfield  {author} {\bibinfo {author} {\bibfnamefont {J.}~\bibnamefont
  {Morgner}}, \bibinfo {author} {\bibfnamefont {B.}~\bibnamefont {Tu}},
  \bibinfo {author} {\bibfnamefont {C.}~\bibnamefont {K\"{o}nig}},\ and\
  \bibinfo {author} {\bibnamefont {et~al.}},\ }\bibfield  {title} {\bibinfo
  {title} {Stringent test of {QED} with hydrogen-like tin},\ }\href
  {https://doi.org/https://doi.org/10.1038/s41586-023-06453-2} {\bibfield
  {journal} {\bibinfo  {journal} {Nature}\ }\textbf {\bibinfo {volume} {622}},\
  \bibinfo {pages} {53} (\bibinfo {year} {2023})}\BibitemShut {NoStop}%
\bibitem [{\citenamefont {Okumura}\ \emph {et~al.}(2023)\citenamefont
  {Okumura}, \citenamefont {Azuma}, \citenamefont {Bennett},\ and\
  \citenamefont {et~al.}}]{PhysRevLett.130.173001}%
  \BibitemOpen
  \bibfield  {author} {\bibinfo {author} {\bibfnamefont {T.}~\bibnamefont
  {Okumura}}, \bibinfo {author} {\bibfnamefont {T.}~\bibnamefont {Azuma}},
  \bibinfo {author} {\bibfnamefont {D.~A.}\ \bibnamefont {Bennett}},\ and\
  \bibinfo {author} {\bibnamefont {et~al.}},\ }\bibfield  {title} {\bibinfo
  {title} {Proof-of-principle experiment for testing strong-field quantum
  electrodynamics with exotic atoms: High precision {X}-ray spectroscopy of
  muonic neon},\ }\href {https://doi.org/10.1103/PhysRevLett.130.173001}
  {\bibfield  {journal} {\bibinfo  {journal} {Phys. Rev. Lett.}\ }\textbf
  {\bibinfo {volume} {130}},\ \bibinfo {pages} {173001} (\bibinfo {year}
  {2023})}\BibitemShut {NoStop}%
\bibitem [{\citenamefont {Antognini}\ \emph {et~al.}(2020)\citenamefont
  {Antognini}, \citenamefont {Berger}, \citenamefont {Cocolios}, \citenamefont
  {Dressler}, \citenamefont {Eichler}, \citenamefont {Eggenberger},
  \citenamefont {Indelicato}, \citenamefont {Jungmann}, \citenamefont {Keitel},
  \citenamefont {Kirch}, \citenamefont {Knecht}, \citenamefont {Michel},
  \citenamefont {Nuber}, \citenamefont {Oreshkina}, \citenamefont {Ouf},
  \citenamefont {Papa}, \citenamefont {Pohl}, \citenamefont {Pospelov},
  \citenamefont {Rapisarda}, \citenamefont {Ritjoho}, \citenamefont {Roccia},
  \citenamefont {Severijns}, \citenamefont {Skawran}, \citenamefont {Vogiatzi},
  \citenamefont {Wauters},\ and\ \citenamefont {Willmann}}]{Antognini}%
  \BibitemOpen
  \bibfield  {author} {\bibinfo {author} {\bibfnamefont {A.}~\bibnamefont
  {Antognini}}, \bibinfo {author} {\bibfnamefont {N.}~\bibnamefont {Berger}},
  \bibinfo {author} {\bibfnamefont {T.~E.}\ \bibnamefont {Cocolios}}, \bibinfo
  {author} {\bibfnamefont {R.}~\bibnamefont {Dressler}}, \bibinfo {author}
  {\bibfnamefont {R.}~\bibnamefont {Eichler}}, \bibinfo {author} {\bibfnamefont
  {A.}~\bibnamefont {Eggenberger}}, \bibinfo {author} {\bibfnamefont
  {P.}~\bibnamefont {Indelicato}}, \bibinfo {author} {\bibfnamefont
  {K.}~\bibnamefont {Jungmann}}, \bibinfo {author} {\bibfnamefont {C.~H.}\
  \bibnamefont {Keitel}}, \bibinfo {author} {\bibfnamefont {K.}~\bibnamefont
  {Kirch}}, \bibinfo {author} {\bibfnamefont {A.}~\bibnamefont {Knecht}},
  \bibinfo {author} {\bibfnamefont {N.}~\bibnamefont {Michel}}, \bibinfo
  {author} {\bibfnamefont {J.}~\bibnamefont {Nuber}}, \bibinfo {author}
  {\bibfnamefont {N.~S.}\ \bibnamefont {Oreshkina}}, \bibinfo {author}
  {\bibfnamefont {A.}~\bibnamefont {Ouf}}, \bibinfo {author} {\bibfnamefont
  {A.}~\bibnamefont {Papa}}, \bibinfo {author} {\bibfnamefont {R.}~\bibnamefont
  {Pohl}}, \bibinfo {author} {\bibfnamefont {M.}~\bibnamefont {Pospelov}},
  \bibinfo {author} {\bibfnamefont {E.}~\bibnamefont {Rapisarda}}, \bibinfo
  {author} {\bibfnamefont {N.}~\bibnamefont {Ritjoho}}, \bibinfo {author}
  {\bibfnamefont {S.}~\bibnamefont {Roccia}}, \bibinfo {author} {\bibfnamefont
  {N.}~\bibnamefont {Severijns}}, \bibinfo {author} {\bibfnamefont
  {A.}~\bibnamefont {Skawran}}, \bibinfo {author} {\bibfnamefont {S.~M.}\
  \bibnamefont {Vogiatzi}}, \bibinfo {author} {\bibfnamefont {F.}~\bibnamefont
  {Wauters}},\ and\ \bibinfo {author} {\bibfnamefont {L.}~\bibnamefont
  {Willmann}},\ }\bibfield  {title} {\bibinfo {title} {Measurement of the
  quadrupole moment of $^{185}\mathrm{Re}$ and $^{187}\mathrm{Re}$ from the
  hyperfine structure of muonic x rays},\ }\href
  {https://doi.org/10.1103/PhysRevC.101.054313} {\bibfield  {journal} {\bibinfo
   {journal} {Phys. Rev. C}\ }\textbf {\bibinfo {volume} {101}},\ \bibinfo
  {pages} {054313} (\bibinfo {year} {2020})}\BibitemShut {NoStop}%
\bibitem [{\citenamefont {Karshenboim}\ \emph {et~al.}(2006)\citenamefont
  {Karshenboim}, \citenamefont {Ivanov},\ and\ \citenamefont
  {Korzinin}}]{Karshenboim2006}%
  \BibitemOpen
  \bibfield  {author} {\bibinfo {author} {\bibfnamefont {S.}~\bibnamefont
  {Karshenboim}}, \bibinfo {author} {\bibfnamefont {V.}~\bibnamefont
  {Ivanov}},\ and\ \bibinfo {author} {\bibfnamefont {E.}~\bibnamefont
  {Korzinin}},\ }\bibfield  {title} {\bibinfo {title} {Study of nuclear
  properties with muonic atoms},\ }\href
  {https://doi.org/https://doi.org/10.1140/epjd/e2006-00133-0} {\bibfield
  {journal} {\bibinfo  {journal} {Eur. Phys. J. D}\ }\textbf {\bibinfo {volume}
  {39}},\ \bibinfo {pages} {351} (\bibinfo {year} {2006})}\BibitemShut
  {NoStop}%
\bibitem [{\citenamefont {Pachucki}\ \emph {et~al.}(2024)\citenamefont
  {Pachucki}, \citenamefont {Lensky}, \citenamefont {Hagelstein}, \citenamefont
  {Li~Muli}, \citenamefont {Bacca},\ and\ \citenamefont
  {Pohl}}]{RevModPhys.96.015001}%
  \BibitemOpen
  \bibfield  {author} {\bibinfo {author} {\bibfnamefont {K.}~\bibnamefont
  {Pachucki}}, \bibinfo {author} {\bibfnamefont {V.}~\bibnamefont {Lensky}},
  \bibinfo {author} {\bibfnamefont {F.}~\bibnamefont {Hagelstein}}, \bibinfo
  {author} {\bibfnamefont {S.~S.}\ \bibnamefont {Li~Muli}}, \bibinfo {author}
  {\bibfnamefont {S.}~\bibnamefont {Bacca}},\ and\ \bibinfo {author}
  {\bibfnamefont {R.}~\bibnamefont {Pohl}},\ }\bibfield  {title} {\bibinfo
  {title} {Comprehensive theory of the {Lamb} shift in light muonic atoms},\
  }\href {https://doi.org/10.1103/RevModPhys.96.015001} {\bibfield  {journal}
  {\bibinfo  {journal} {Rev. Mod. Phys.}\ }\textbf {\bibinfo {volume} {96}},\
  \bibinfo {pages} {015001} (\bibinfo {year} {2024})}\BibitemShut {NoStop}%
\bibitem [{\citenamefont {Borie}(2012)}]{BORIE2012733}%
  \BibitemOpen
  \bibfield  {author} {\bibinfo {author} {\bibfnamefont {E.}~\bibnamefont
  {Borie}},\ }\bibfield  {title} {\bibinfo {title} {Lamb shift in light muonic
  atoms -- {Revisited}},\ }\href
  {https://doi.org/https://doi.org/10.1016/j.aop.2011.11.017} {\bibfield
  {journal} {\bibinfo  {journal} {Annals of Physics}\ }\textbf {\bibinfo
  {volume} {327}},\ \bibinfo {pages} {733} (\bibinfo {year}
  {2012})}\BibitemShut {NoStop}%
\bibitem [{\citenamefont {Chen}(1970)}]{PhysRevC.1.1176}%
  \BibitemOpen
  \bibfield  {author} {\bibinfo {author} {\bibfnamefont {M.}~\bibnamefont
  {Chen}},\ }\bibfield  {title} {\bibinfo {title} {Nuclear polarization in
  muonic atoms of deformed nuclei},\ }\href
  {https://doi.org/10.1103/PhysRevC.1.1176} {\bibfield  {journal} {\bibinfo
  {journal} {Phys. Rev. C}\ }\textbf {\bibinfo {volume} {1}},\ \bibinfo {pages}
  {1176} (\bibinfo {year} {1970})}\BibitemShut {NoStop}%
\bibitem [{\citenamefont {Valuev}\ and\ \citenamefont
  {Oreshkina}(2024)}]{PhysRevA.109.042811}%
  \BibitemOpen
  \bibfield  {author} {\bibinfo {author} {\bibfnamefont {I.~A.}\ \bibnamefont
  {Valuev}}\ and\ \bibinfo {author} {\bibfnamefont {N.~S.}\ \bibnamefont
  {Oreshkina}},\ }\bibfield  {title} {\bibinfo {title} {Full leading-order
  nuclear polarization in highly charged ions},\ }\href
  {https://doi.org/10.1103/PhysRevA.109.042811} {\bibfield  {journal} {\bibinfo
   {journal} {Phys. Rev. A}\ }\textbf {\bibinfo {volume} {109}},\ \bibinfo
  {pages} {042811} (\bibinfo {year} {2024})}\BibitemShut {NoStop}%
\bibitem [{\citenamefont {Borie}\ and\ \citenamefont
  {Rinker}(1978)}]{PhysRevA.18.324}%
  \BibitemOpen
  \bibfield  {author} {\bibinfo {author} {\bibfnamefont {E.}~\bibnamefont
  {Borie}}\ and\ \bibinfo {author} {\bibfnamefont {G.~A.}\ \bibnamefont
  {Rinker}},\ }\bibfield  {title} {\bibinfo {title} {Improved calculation of
  the muonic-helium lamb shift},\ }\href
  {https://doi.org/10.1103/PhysRevA.18.324} {\bibfield  {journal} {\bibinfo
  {journal} {Phys. Rev. A}\ }\textbf {\bibinfo {volume} {18}},\ \bibinfo
  {pages} {324} (\bibinfo {year} {1978})}\BibitemShut {NoStop}%
\bibitem [{\citenamefont {Yerokhin}\ and\ \citenamefont
  {Oreshkina}(2023)}]{PhysRevA.108.052824}%
  \BibitemOpen
  \bibfield  {author} {\bibinfo {author} {\bibfnamefont {V.~A.}\ \bibnamefont
  {Yerokhin}}\ and\ \bibinfo {author} {\bibfnamefont {N.~S.}\ \bibnamefont
  {Oreshkina}},\ }\bibfield  {title} {\bibinfo {title} {{QED} calculations of
  the nuclear recoil effect in muonic atoms},\ }\href
  {https://doi.org/10.1103/PhysRevA.108.052824} {\bibfield  {journal} {\bibinfo
   {journal} {Phys. Rev. A}\ }\textbf {\bibinfo {volume} {108}},\ \bibinfo
  {pages} {052824} (\bibinfo {year} {2023})}\BibitemShut {NoStop}%
\bibitem [{\citenamefont {Oreshkina}(2022)}]{PhysRevResearch.4.L042040}%
  \BibitemOpen
  \bibfield  {author} {\bibinfo {author} {\bibfnamefont {N.~S.}\ \bibnamefont
  {Oreshkina}},\ }\bibfield  {title} {\bibinfo {title} {Self-energy correction
  to the energy levels of heavy muonic atoms},\ }\href
  {https://doi.org/10.1103/PhysRevResearch.4.L042040} {\bibfield  {journal}
  {\bibinfo  {journal} {Phys. Rev. Res.}\ }\textbf {\bibinfo {volume} {4}},\
  \bibinfo {pages} {L042040} (\bibinfo {year} {2022})}\BibitemShut {NoStop}%
\bibitem [{\citenamefont {Borie}\ and\ \citenamefont
  {Rinker}(1982)}]{RevModPhys.54.67}%
  \BibitemOpen
  \bibfield  {author} {\bibinfo {author} {\bibfnamefont {E.}~\bibnamefont
  {Borie}}\ and\ \bibinfo {author} {\bibfnamefont {G.~A.}\ \bibnamefont
  {Rinker}},\ }\bibfield  {title} {\bibinfo {title} {The energy levels of
  muonic atoms},\ }\href {https://doi.org/10.1103/RevModPhys.54.67} {\bibfield
  {journal} {\bibinfo  {journal} {Rev. Mod. Phys.}\ }\textbf {\bibinfo {volume}
  {54}},\ \bibinfo {pages} {67} (\bibinfo {year} {1982})}\BibitemShut {NoStop}%
\bibitem [{\citenamefont {Breidenbach}\ \emph {et~al.}(2022)\citenamefont
  {Breidenbach}, \citenamefont {Dizer}, \citenamefont {Cakir},\ and\
  \citenamefont {Harman}}]{Breidenbach_PhysRevA.106.042805}%
  \BibitemOpen
  \bibfield  {author} {\bibinfo {author} {\bibfnamefont {S.}~\bibnamefont
  {Breidenbach}}, \bibinfo {author} {\bibfnamefont {E.}~\bibnamefont {Dizer}},
  \bibinfo {author} {\bibfnamefont {H.}~\bibnamefont {Cakir}},\ and\ \bibinfo
  {author} {\bibfnamefont {Z.}~\bibnamefont {Harman}},\ }\bibfield  {title}
  {\bibinfo {title} {Hadronic vacuum polarization correction to atomic energy
  levels},\ }\href {https://doi.org/10.1103/PhysRevA.106.042805} {\bibfield
  {journal} {\bibinfo  {journal} {Phys. Rev. A}\ }\textbf {\bibinfo {volume}
  {106}},\ \bibinfo {pages} {042805} (\bibinfo {year} {2022})}\BibitemShut
  {NoStop}%
\bibitem [{\citenamefont {Burkhardt}\ \emph {et~al.}(1989)\citenamefont
  {Burkhardt}, \citenamefont {Jegerlehner}, \citenamefont {Penso},\ and\
  \citenamefont {et~al.}}]{BURKHARDT1989}%
  \BibitemOpen
  \bibfield  {author} {\bibinfo {author} {\bibfnamefont {H.}~\bibnamefont
  {Burkhardt}}, \bibinfo {author} {\bibfnamefont {F.}~\bibnamefont
  {Jegerlehner}}, \bibinfo {author} {\bibfnamefont {G.}~\bibnamefont {Penso}},\
  and\ \bibinfo {author} {\bibnamefont {et~al.}},\ }\bibfield  {title}
  {\bibinfo {title} {Uncertainties in the hadronic contribution to the qed
  vacuum polarization},\ }\href
  {https://doi.org/https://doi.org/10.1007/BF01506546} {\bibfield  {journal}
  {\bibinfo  {journal} {Z. Phys. C - Particles and Fields}\ }\textbf {\bibinfo
  {volume} {43}},\ \bibinfo {pages} {497} (\bibinfo {year} {1989})}\BibitemShut
  {NoStop}%
\bibitem [{\citenamefont {Sundaresan}\ and\ \citenamefont
  {Watson}(1975)}]{PhysRevD.11.230}%
  \BibitemOpen
  \bibfield  {author} {\bibinfo {author} {\bibfnamefont {M.~K.}\ \bibnamefont
  {Sundaresan}}\ and\ \bibinfo {author} {\bibfnamefont {P.~J.~S.}\ \bibnamefont
  {Watson}},\ }\bibfield  {title} {\bibinfo {title} {Hadronic
  vacuum-polarization corrections in muonic atoms},\ }\href
  {https://doi.org/10.1103/PhysRevD.11.230} {\bibfield  {journal} {\bibinfo
  {journal} {Phys. Rev. D}\ }\textbf {\bibinfo {volume} {11}},\ \bibinfo
  {pages} {230} (\bibinfo {year} {1975})}\BibitemShut {NoStop}%
\bibitem [{\citenamefont {Burkhardt}\ and\ \citenamefont
  {Pietrzyk}(2001)}]{BURKHARDT2001}%
  \BibitemOpen
  \bibfield  {author} {\bibinfo {author} {\bibfnamefont {H.}~\bibnamefont
  {Burkhardt}}\ and\ \bibinfo {author} {\bibfnamefont {B.}~\bibnamefont
  {Pietrzyk}},\ }\bibfield  {title} {\bibinfo {title} {Update of the hadronic
  contribution to the qed vacuum polarization},\ }\href
  {https://doi.org/https://doi.org/10.1016/S0370-2693(01)00393-8} {\bibfield
  {journal} {\bibinfo  {journal} {Physics Letters B}\ }\textbf {\bibinfo
  {volume} {513}},\ \bibinfo {pages} {46} (\bibinfo {year} {2001})}\BibitemShut
  {NoStop}%
\bibitem [{\citenamefont {Angeli}\ and\ \citenamefont
  {Marinova}(2013)}]{Angeli2013Table}%
  \BibitemOpen
  \bibfield  {author} {\bibinfo {author} {\bibfnamefont {I.}~\bibnamefont
  {Angeli}}\ and\ \bibinfo {author} {\bibfnamefont {K.}~\bibnamefont
  {Marinova}},\ }\bibfield  {title} {\bibinfo {title} {Table of experimental
  nuclear ground state charge radii: An update},\ }\href
  {https://doi.org/https://doi.org/10.1016/j.adt.2011.12.006} {\bibfield
  {journal} {\bibinfo  {journal} {Atomic Data and Nuclear Data Tables}\
  }\textbf {\bibinfo {volume} {99}},\ \bibinfo {pages} {69} (\bibinfo {year}
  {2013})}\BibitemShut {NoStop}%
\bibitem [{\citenamefont {Friar}\ and\ \citenamefont
  {Martorell}(1999)}]{Friar}%
  \BibitemOpen
  \bibfield  {author} {\bibinfo {author} {\bibfnamefont {J.~L.}\ \bibnamefont
  {Friar}}\ and\ \bibinfo {author} {\bibfnamefont {J.}~\bibnamefont
  {Martorell}},\ }\bibfield  {title} {\bibinfo {title} {Hadronic vacuum
  polarization and the lamb shift},\ }\href
  {https://doi.org/https://doi.org/10.1103/PhysRevA.59.4061} {\bibfield
  {journal} {\bibinfo  {journal} {Phys. Rev. A}\ }\textbf {\bibinfo {volume}
  {59}},\ \bibinfo {pages} {4061} (\bibinfo {year} {1999})}\BibitemShut
  {NoStop}%
\end{thebibliography}%
\end{document}